\documentclass[journal, a4paper, 9pt, onecolumn]{article}

\usepackage{spconf}\ninept
\usepackage[T1]{fontenc} 
\usepackage[utf8]{inputenc}

\usepackage{microtype}
\usepackage{amsmath,amssymb,mathtools}

\usepackage[noadjust]{cite}
\usepackage{subcaption}

\usepackage{url}

\usepackage{siunitx}
\usepackage{units}
\usepackage{pgfplots}
\newlength\figureheight
\newlength\figurewidth 
\pgfplotsset{compat=newest}
\pgfplotsset{plot coordinates/math parser=false} 
\pgfplotsset{every x tick label/.append style={font=\scriptsize, yshift=0ex}}
\pgfplotsset{every y tick label/.append style={font=\scriptsize, xshift=0ex}}
\pgfplotsset{every axis legend/.append style={font=\scriptsize}}

\usepackage{xspace}
\usepackage{amsmath}
\usepackage{scalerel}
\usepackage{xcolor}

\newcommand{\forkortning}[1]{\relax\ifmmode \scaleobj{0.888}{\mathsf{\MakeUppercase{#1}}} \else \textsc{#1}\xspace\fi}

\newcommand{\ADC}{\textsc{adc}\xspace}
\newcommand{\ADCs}{\textsc{adc}s\xspace}
\newcommand{\AGC}{\textsc{agc}\xspace}

\newcommand{\MIMO}{\textsc{mimo}\xspace}

\newcommand{\MSE}{\textsc{mse}\xspace}

\newcommand{\SNR}{\forkortning{snr}}

\newcommand{\MIMOACLR}{\relax\ifmmode\scaleobj{0.888}{\mathsf{MIMO}\text{-}\mathsf{ACLR}}\else\textsc{mimo-aclr}\xspace\fi}

\newcommand{\Exp}{\ensuremath{\operatorname{\mathsf{E}}}}
\DeclareMathOperator*{\argmin}{arg\,min}

\def\Re{\operatorname{\mathfrak{Re}}}
\def\Im{\operatorname{\mathfrak{Im}}}

\newcounter{run}
\usepackage{catchfile}
\IfFileExists{\jobname.runs}{%
	\begingroup
	\newcommand{\tmp}{\input{\jobname.runs}}%
	\setcounter{run}{\tmp}%
	\endgroup
}{}
\stepcounter{run}

\usepackage{atveryend}
\usepackage{newfile}
\AtVeryEndDocument{%
	\newoutputstream{runs}%
	\openoutputfile{\jobname.runs}{runs}%
	\addtostream{runs}{\number\value{run}}%
	\closeoutputstream{runs}%
}

\newsavebox{\mybox}

\definecolor{color1}{HTML}{00B4E7}%
\definecolor{color2}{HTML}{FF431A}%
\definecolor{color3}{HTML}{002F3D}%
\definecolor{color4}{HTML}{7300E6}%
\definecolor{color5}{HTML}{228B22}%

\makeatletter
\@ifpackageloaded{unicode-math}{}{%
	\let\symcal\mathcal%
	\DeclareMathAlphabet{\mathsfit}{\encodingdefault}{\sfdefault}{m}{sl}
	\DeclareMathAlphabet{\mathbfit}{\encodingdefault}{\rmdefault}{b}{it}
	
	\let\symsfit\mathsfit
	\DeclareMathAlphabet{\symbfsf}{\encodingdefault}{\sfdefault}{bx}{n}
	\let\symbb\mathbb%
}
\makeatother

\newcommand{\CN}{\operatorname{\symcal{C}\kern-.16em\symcal{N}}}

\newcommand{\jia}{\textsc{liu}}%
\newcommand{\yi}{\textsc{postech}}%
\newcommand{\bing}{\textsc{uta}}%

\begin{document}

	\title{Achievable Uplink Rates for Massive MIMO with Coarse Quantization}

	\name{Christopher~Mollén\nobreak\hspace{.05em}$^{\text{\jia}}$, Junil~Choi\nobreak\hspace{.05em}$^{\text{\yi}}$, Erik~G.~Larsson\nobreak\hspace{.05em}$^{\text{\jia}}$, and~Robert~W.~Heath~Jr.$^{\text{\bing}}$%
		}
	
	\address{$^{\text{\jia}}$ Linköping University, Dept.\ of Electrical Engineering, 581 83 Linköping, Sweden \\
		$^{\text{\yi}}$  POSTECH, Dept.\ of Electrical Engineering, Pohang 37673, South Korea\\
		\clap{$^{\text{\bing}}$ University of Texas at Austin, Dept.\ of Electrical and Computer Engineering, Austin, TX 78712, USA}}

	\maketitle
	\begin{abstract}
		The high hardware complexity of a massive \MIMO base station, which requires hundreds of radio chains, makes it challenging to build commercially.  One way to reduce the hardware complexity and power consumption of the receiver is to lower the resolution of the analog-to-digital converters (\ADCs).  We derive an achievable rate for a massive \MIMO system with arbitrary quantization and use this rate to show that \ADCs with as low as \unit[3]{bits} can be used without significant performance loss at spectral efficiencies around \unit[3.5]{bpcu} per user, also under interference from stronger transmitters and with some imperfections in the automatic gain control.
	\end{abstract}

\def\Re{{\operatorname{\mathfrak{Re}}}}
\def\Im{{\operatorname{\mathfrak{Im}}}}
	
\section{Introduction}
Massive \MIMO is a promising technology for the improvement of today's wireless infrastructure \cite{6690}.  The huge number of transceiver chains required in massive \MIMO base stations, however, makes their hardware complexity and cost a challenge that has to be overcome before the technology can become commercially viable \cite{athley2015analysisEUCAP}.  It has been proposed to build each transceiver chain from low-end hardware to reduce the complexity \cite{bjornson2013massive}.  

In this paper, we perform an information theoretical analysis of a massive \MIMO system with arbitrary \ADCs and derive an achievable rate, which takes quantization into account, for a linear combiner that uses low-complexity channel estimation.  The achievable rate is used to draw the conclusion that \ADCs with \unit[3]{bits} are sufficient to achieve a rate close to that of an unquantized system, see Section~\ref{sec:conclusions} for more detailed conclusions.  This analysis is an extension of work in \cite{mollen2016uplinkArXiv}, where we only study one-bit \ADCs.

Previous work has studied the capacity of the one-bit quantized frequency-flat \MIMO channel \cite{mezghani2007ultra, Mo_Jianhua_TSP15}, developed detection and channel estimation methods for the frequency-flat multiuser \MIMO channel \cite{ivrlac2007mimo, choi2016nearTCOM, risi2014massive} and for the frequency-selective channel \cite{dabeer2010channel, wang2015multiuser}.  Low-resolution \ADCs were studied in \cite{studer2016quantized} and the use of a mix of \ADCs with different resolutions in \cite{liang2015mixed}.  While the methods for frequency-flat channels are hard to extend to frequency-selective channels and the methods for frequency-selective channels either have high computational complexity, require long pilot sequences or imply impractical design changes to the massive \MIMO base station, the linear detector and channel estimator that we study is the same low-complexity methods that has been proven possible to implement in practical testbeds \cite{shepard2012argos, vieira2014flexible}.

A parametric model for hardware imperfections was proposed in \cite{bjornson2015massive}, where the use of low-resolution \ADCs in massive \MIMO also was suggested.  The parametric model is used in \cite{verenzuela2016hardware} to show that \unit[4--5]{bits} of resolution maximizes the spectral efficiency for a given power consumption.  Several system simulations have been performed to analyze low-resolution \ADCs, e.g. \cite{jacobsson2016massive, desset2015validation}, where the conclusions coincide with the conclusions in this paper: that three-bit \ADCs are sufficient to obtain a performance close to an unquantized system.

\section{System Model}
The uplink transmission from $K$ single-antenna users to a massive \MIMO base station with $M$ antennas is studied.  The transmission is based on pulse-amplitude modulation and, for the reception, a matched filter is used for demodulation.  It is assumed that the matched filter is implemented as an analog filter and that its output is sampled at symbol rate by an \ADC with finite resolution.  Because the nonlinear quantization of the \ADC comes after the matched filter, the transmission can be studied in symbol-sampled discrete time.

Each user $k$ transmits the signal $\sqrt{P_k}x_k[n]$, which is normalized,
\begin{align}
	\Exp\bigl[|x_k[n]|^2\bigr] = 1,
\end{align}
so that $P_k$ denotes the transmit power.  The channel from user $k$ to antenna $m$ at the base station is described by its impulse response $\sqrt{\beta_k} h_{mk}[\ell]$, which can be factorized into a large-scale fading coefficient $\beta_k$ and a small-scale fading impulse response $h_{mk}[\ell]$.  The large-scale fading varies slowly in comparison to the symbol rate and can be accurately estimated with little overhead by both user and base station.  It is therefore assumed to be known throughout the system.  The small-scale fading, in contrast, is \textit{a priori} unknown to everybody.  It is independent across $\ell$ and follows the power delay profile
\begin{align}
\sigma^2_k[\ell] \triangleq \Exp\left[|h_{mk}[\ell]|^2\right],
\end{align}
however, is assumed to be known.  It is also assumed that $\sigma^2_k[\ell] = 0$ for all $\ell \notin [0,\ldots,L{-}1]$.  Since variations in received power should be described by the large-scale fading only, the power delay profile is normalized such that
\begin{align}
	\sum_{\ell=0}^{L-1} \sigma^2_k[\ell] = 1, \quad\forall k.
\end{align}
Base station antenna $m$ receives the signal
\begin{align}
	y_m[n] = \sum_{k=1}^{K} \sqrt{\beta_k P_k} \sum_{\ell=0}^{L-1} h_{mk}[\ell] x_k[n - \ell] + z_m[n].
\end{align}
The thermal noise of the receiver $z_m[n]$ is modeled as a white stochastic process, for which $z_m[n] \sim \CN(0,N_0)$.  The received power is denoted
\begin{align}
	P_\text{rx} \triangleq \Exp \left[ |y_m[n]|^2 \right] = \sum_{k=1}^{K} \beta_k P_k + N_0.
\end{align}

Transmission is assumed to be done with a cyclic prefix in blocks of $N$ symbols.  The received signal can than be given in the frequency domain as
\begin{align}\label{eq:FT_sig}
	\symsfit{y}_m[\nu] \triangleq \smash[b]{\frac{1}{\sqrt{N}} \sum_{n=0}^{N-1} y_m[n] e^{-j2\pi n \nu / N} = \sum_{k=1}^{K} \symsfit{h}_{mk}[\nu] \symsfit{x}_k[\nu] + \symsfit{z}_m[\nu]},
\end{align}
The Fourier transforms $\symsfit{x}_k[\nu]$ and $\symsfit{z}_k[\nu]$ of the transmit signal $x_k[n]$ and noise $z_m[n]$ are defined in the same way as $\symsfit{y}_m[\nu]$.  The frequency response of the channel is defined as
\begin{align}
	\symsfit{h}_{mk}[\nu] \triangleq \sum_{\ell=0}^{L-1} h_{mk}[\ell] e^{-j2\pi \ell\nu/N}.
\end{align}

\section{Quantization}
The inphase and quadrature signals are assumed to be quantized separately by two identical \ADCs with quantization levels given by $\symcal{Q}_{\Re} \subseteq \symbb{R}$.  The set of quantization points is denoted $\symcal{Q} \triangleq \{a+jb : a,b \in \symcal{Q}_{\Re}\}$ and the quantization by
\begin{align}
	\left[y\right]_\symcal{Q} \triangleq \argmin_{q \in \symcal{Q}} \left|y - q \right|.
\end{align}
To adjust the input signal to the dynamic range of the \ADC, an automatic gain control scales the input power by $A$.  The \ADC outputs:
\begin{align}
	q_m[n] \triangleq  \left[ \sqrt{A} y_m[n] \right]_\symcal{Q}.
\end{align}
To analyze the effect of the quantization, the quantized signal is partitioned into one part $\rho y_m[n]$ that is correlated to the transmit signal and one part $e_m[n]$ that is uncorrelated:
\begin{align}\label{eq:quant_error}
q_m[n] &= \rho y_m[n] + e_m[n]
\end{align}
where the constant $\rho$ and the variance of the uncorrelated part can be derived through the orthogonality principle:
\begin{align}
\rho &= \frac{\Exp\left[q_m[n] y^*_m[n]\right]}{\Exp\left[|y_m[n]|^2\right]},\\
\Exp\left[|e_m[n]|^2\right] &= \Exp\left[|q_m[n]|^2\right] - \frac{\Bigl|\Exp\left[q_m[n] y^*_m[n]\right]\Bigr|^2}{\Exp\left[|y_m[n]|^2\right]}.
\end{align}
The normalized mean-square error (\MSE) of the quantization is denoted by
\begin{align}
	Q &\triangleq \frac{1}{|\rho|^2} \Exp\left[|e_m[n]|^2\right]\\
	&= P_\text{rx} \left( \frac{\Exp\left[|q_m[n]|^2\right] \Exp\left[|y_m[n]|^2\right]}{\Bigl|\Exp\left[q_m[n] y^*_m[n]\right] \Bigr|^2} - 1 \right).
\end{align}

\begin{figure}
	\begin{tikzpicture}
	\def\xmin{-10}
	\def\xmax{10}
	\begin{axis}[
	xmin=\xmin, 
	xmax=\xmax, 
	ymin=0,
	ymax=0.19,
	legend pos = south east, 
	height=0.85\figureheight,
	width=0.85\linewidth,
	xlabel={\AGC imperfection $A / A^\star$ [dB]},
	ylabel={quantization \textsc{mse}\\ $Q / P_\text{rx}$},
	ylabel style={align=center},
	ymajorgrids,
	ytick = {0.5708, 0.1331, 0.03576, 0.009573},
	yticklabels={0.5708, 0.1331, 0.03576, 0.009573},
	]
	\addplot[mark=none, color=color3] table [row sep=\\] {
		-10 0.20966\\
		-9 0.16816\\
		-8 0.13523\\
		-7 0.10915\\
		-6 0.088517\\
		-5 0.072269\\
		-4 0.059519\\
		-3 0.049603\\
		-2 0.042234\\
		-1 0.037484\\
		0 0.03576\\
		1 0.037651\\
		2 0.043514\\
		3 0.053474\\
		4 0.067353\\
		5 0.084674\\
		6 0.10493\\
		7 0.12742\\
		8 0.15153\\
		9 0.17665\\
		10 0.20223\\
	};
	\node[anchor=south west, font=\scriptsize\color{color3}] at (axis cs:-5.8, 0.075) {three-bit \ADC};
	
	\addplot[mark=none, color=color4] table [row sep=\\] {
		-10 0.057047\\
		-9 0.045752\\
		-8 0.036781\\
		-7 0.029675\\
		-6 0.024055\\
		-5 0.019614\\
		-4 0.016148\\
		-3 0.013505\\
		-2 0.011498\\
		-1 0.010142\\
		0 0.009573\\
		1 0.010330\\
		2 0.012964\\
		3 0.018226\\
		4 0.026642\\
		5 0.038561\\
		6 0.053913\\
		7 0.072405\\
		8 0.093582\\
		9 0.116857\\
		10 0.14168\\
		};
	\node[anchor=south west, font=\scriptsize\color{color4}] at (axis cs:-9.1, 0.04) {four-bit \ADC};
	\end{axis}
	\end{tikzpicture}
	\caption{Quantization \MSE for optimal four-bit \ADC with imperfect \AGC.}
	\label{fig:imperf_AGC}
\end{figure}
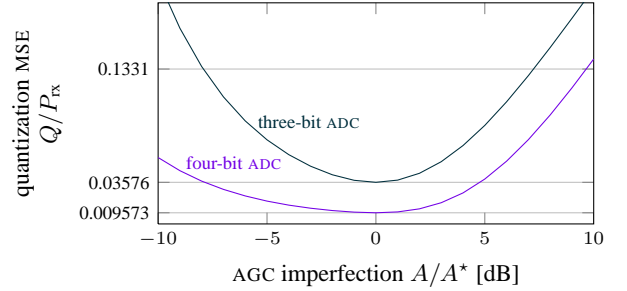

An \ADC with $b$\mbox{-}bit resolution has $|\symcal{Q}_\Re| = 2^b$ quantization levels.  In \cite{max1960quantizing}, the quantization levels that minimize the \MSE for a Gaussian input signal with unit variance are derived numerically for \unit[1--5]{bit} \ADCs, both with arbitrarily and uniformly spaced quantization levels.  The normalized \MSE of the quantization has been computed numerically and is given in Table~\ref{tab:QMSE} for the optimized quantizers.  To obtain the \MSE in Table~\ref{tab:QMSE} with the quantization levels from \cite{max1960quantizing}, the input power has to be unity and the automatic gain control $A = A^\star \triangleq 1 / P_\text{rx}$.  Figure~\ref{fig:imperf_AGC} shows how the quantization \MSE in a four-bit \ADC changes with imperfect gain control.  Even if the gain control varies between $-$8 and \unit[5]{dB} from the optimal value, the \MSE is still better than that of a three-bit \ADC.

\def\onebiterror{0.5708}
\def\twobiterror{0.1331}
\def\threebiterror{0.03576}
\def\fourbiterror{0.009573}
\def\fivebiterror{0.002492}

\begin{table}\footnotesize\setlength{\tabcolsep}{4pt}\centering
	\caption{Normalized quantization mean square-error $Q / P_\text{rx}$}
	\begin{tabular}{l|lllll}
		resolution $b$ & 1 & 2 & 3 & 4 & 5\\
		\hline
		\rule{0pt}{2.6ex}%
		optimal levels & 0.5708 & 0.1331 & 0.03576 & 0.009573 & 0.002492\\
		uniform levels & 0.5708 & 0.1349 & 0.03889 & 0.01166 & 0.003506\\
	\end{tabular}
	\label{tab:QMSE}
\end{table}

\section{Channel Estimation}
Channel estimation is done by receiving $N = N_\text{p}$\mbox{-}symbol long orthogonal pilots from the users, i.e., pilots $x_k[n]$ such that:
\begin{align}\label{eq:pilot_requirement}
\sum_{n=0}^{N_\text{p} - 1} x_k[n] x_{k'}^*[n+\ell] = 
\begin{cases}
	N_\text{p}, &\text{if }k=k',\ell = 0\\
	0, &\text{if }k\neq k',\ell=1,\ldots,L-1\\
\end{cases},
\end{align}
where the indices are taken modulo $N_\text{p}$.  To fulfill \eqref{eq:pilot_requirement}, $N_\text{p} \geq KL$.  We will call the factor of extra pilots $\mu \triangleq N_\text{p} / (KL)$ the \emph{pilot excess factor}.  As remarked upon in \cite{mollen2016uplinkArXiv}, not all sequences fulfilling \eqref{eq:pilot_requirement} result in the same performance.  Here we use the pilots proposed in \cite{mollen2016uplinkArXiv}.  Using \eqref{eq:quant_error} and \eqref{eq:pilot_requirement}, an observation of the channel is obtained by correlation:
\begin{align}
	r_{mk}[\ell] &= \frac{1}{\rho\sqrt{N_\text{p}}} \sum_{n=0}^{N_\text{p} - 1} q_m[n] x_k^*[n+\ell]\\
	&= \sqrt{\beta_k P_k N_\text{p}} h_{mk}[\ell] + e'_{mk}[\ell] + z'_{mk}[\ell],\\
	\intertext{where}
	e'_{mk}[\ell] &\triangleq \frac{1}{\rho\sqrt{N_\text{p}}} \sum_{n=0}^{N_\text{p} - 1} e_m[n] x_k^*[n+\ell],\\
	z'_{mk}[\ell] &\triangleq \frac{1}{\sqrt{N_\text{p}}} \sum_{n=0}^{N_\text{p} - 1} z_m[n] x_k^*[n+\ell] \sim \CN\left( 0, N_0 \right).
\end{align}

\begin{figure}[t!]\centering
	\begin{tikzpicture}
	\def\xmin{-10}
	\def\xmax{10}
	\begin{axis}[
	xmin=\xmin, 
	xmax=\xmax, 
	legend pos = south east, 
	height=\figureheight, 
	width=\linewidth,
	xlabel={\SNR $\beta_k P_k / N_0$ [dB]},
	ylabel={\llap{channel} estimation variance $c_k$ \rlap{[dB]}},
	legend cell align=left,
	]
	\def\power{(10^(x/10))}
	\def\nrUsers{5}
	\def\pilotExcessFactor{1}
	
	\addplot[color=black, domain=\xmin:\xmax] {10*ln(\power * \pilotExcessFactor * \nrUsers / (\power * \pilotExcessFactor * \nrUsers + 1))/ln(10)};
	
	\addplot[color=color3, domain=\xmin:\xmax] {10*ln(\power * \pilotExcessFactor * \nrUsers / (\power * \pilotExcessFactor * \nrUsers + (\nrUsers * \power + 1) * \threebiterror + 1))/ln(10)};
	
	\addplot[color=color2, domain=\xmin:\xmax] {10*ln(\power * \pilotExcessFactor * \nrUsers / (\power * \pilotExcessFactor * \nrUsers + (\nrUsers * \power + 1) * \twobiterror + 1))/ln(10)};
	
	\addplot[color=color1, domain=\xmin:\xmax] {10*ln(\power * \pilotExcessFactor * \nrUsers / (\power * \pilotExcessFactor * \nrUsers + (\nrUsers * \power + 1) * \onebiterror + 1))/ln(10)};
	
	\legend{
		no quantization,
		three-bit \ADCs, 
		two-bit \ADCs, 
		one-bit \ADCs}
	\node[anchor=north west,draw,outer sep=2mm] at (rel axis cs:0,1) {$\mu = 1$};
	\end{axis}
	\end{tikzpicture}
	
	\vspace{2ex}
	
	\begin{tikzpicture}
	\def\xmin{1}
	\def\xmax{5}
	\begin{axis}[xmin=\xmin, xmax=\xmax, legend pos = south east, height=\figureheight, width=\linewidth,xlabel={pilot excess factor $\mu = N_\text{p} / (KL)$},ylabel={\llap{channel} estimation variance $c_k$ \rlap{[dB]}},legend cell align=left,samples at = {1,2,3,4,5,6,7,8,9,10,11,12,13,14,15,16}]
	\def\power{(10^(-10/10))}
	\def\nrUsers{5}
	\def\pilotExcessFactor{x}
	
	\addplot[color=black, mark=*, fill = white, mark size = 3pt, domain=\xmin:\xmax] {10*ln(\power * \pilotExcessFactor * \nrUsers / (\power * \pilotExcessFactor * \nrUsers + 1))/ln(10)};
	
	\addplot[color=color3, mark=diamond*, domain=\xmin:\xmax] {10*ln(\power * \pilotExcessFactor * \nrUsers / (\power * \pilotExcessFactor * \nrUsers + (\nrUsers * \power + 1) * \threebiterror + 1))/ln(10)};
	
	\addplot[color=color2, mark=square*, domain=\xmin:\xmax] {10*ln(\power * \pilotExcessFactor * \nrUsers / (\power * \pilotExcessFactor * \nrUsers + (\nrUsers * \power + 1) * \twobiterror + 1))/ln(10)};
	
	\addplot[color=color1, mark=x, domain=\xmin:\xmax] {10*ln(\power * \pilotExcessFactor * \nrUsers / (\power * \pilotExcessFactor * \nrUsers + (\nrUsers * \power + 1) * \onebiterror + 1))/ln(10)};
	
	\legend{no quantization, three-bit \ADCs, two-bit \ADCs, one-bit \ADCs}
	\node[anchor=north west, draw,outer sep=2mm] at (rel axis cs:0,1) {$\beta_k P_k / N_0 = \unit[-10]{dB}$};
	\end{axis}
	\end{tikzpicture}
	\caption{The channel estimation variance with 5 users and a uniform power delay profile $\sigma_k^2[\ell] = 1/L$, for all $k, \ell$, and with equal received power from all users $\beta_k P_k = \beta_1 P_1$, for all $k$.  The optimal quantization levels derived in \cite{max1960quantizing} are used.  Only integer pilot excess factors are considered.}
	\label{fig:channel_quality}
\end{figure}
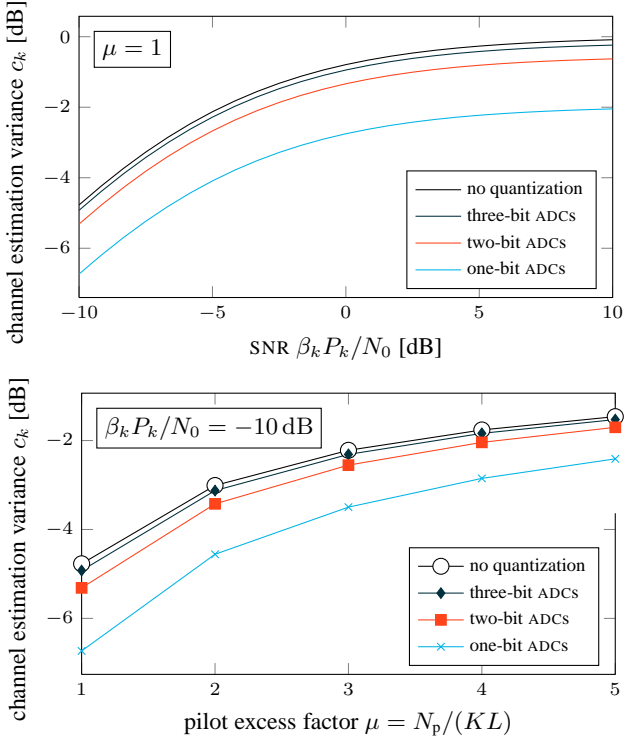

The linear minimum \MSE estimate of the frequency response of the channel is thus
\begin{align}
	\hat{\symsfit{h}}_{mk}[\nu] = \sum_{\ell=0}^{L-1} \frac{\sqrt{\beta_k P_k} \sigma_k^2[\ell]}{\beta_k P_k N_\text{p} \sigma_k^2[\ell] + Q + N_0} r_{mk}[\ell] e^{-j2\pi\ell\nu / N}
\end{align}
and the error $\epsilon_{mk}[\nu] \triangleq \hat{\symsfit{h}}_{mk}[\nu] - \symsfit{h}_{mk}[\nu]$ has the variance $1 - c_k$, where the channel estimation variance is given by
\begin{align}\label{eq:channel_quality}
	c_k \triangleq \Exp\left[|\hat{\symsfit{h}}_{mk}[\nu]|^2\right] = \sum_{\ell=0}^{L-1} \frac{\sigma_k^4[\ell] \beta_k P_k N_\text{p}}{\sigma_{k}^2[\ell] \beta_k P_k N_\text{p} + Q + N_0}.
\end{align}
Figure~\ref{fig:channel_quality} shows the channel estimation variance.  A resolution of \unit[2]{bit} is enough to obtain a channel estimation variance only \unit[0.5]{dB} worse than in an unquantized system.  With a resolution of \unit[3]{bit} or higher, the channel estimation variance is practically the same as that of the unquantized system.  Increasing the pilot length, increases the channel estimation variance in all systems.  The improvement is, however, the largest when going from $\mu=1$ to $\mu=2$; thereafter the improvement gets smaller.

\section{Data Transmission}
The uplink data is transmitted in a block of length $N = N_\text{u}$, which is separated from the pilot block in time.  The received signal is processed in a linear combiner and an estimate of the transmitted signal is obtained by
\begin{align}\label{eq:tx_signal_estimate}
	\hat{\symsfit{x}}_k[\nu] = \frac{1}{\rho} \sum_{m=1}^{M} \symsfit{w}_{mk}[\nu] \symsfit{q}_m[\nu],
\end{align}
where the Fourier transform $\symsfit{q}_m[\nu]$ of $q_m[n]$ is defined in the same way as $\symsfit{y}_m[\nu]$ in \eqref{eq:FT_sig} and the combiner weights $\symsfit{w}_{mk}[\nu]$ are chosen as a function of the channel estimate.  For example, the maximum-ratio and zero-forcing combiners, see \cite{mollen2016uplinkArXiv}, can be used.

If we code over many channel realizations, an achievable rate, independent of $\nu$, is given by \cite{mollen2016uplinkArXiv}:
\begin{align}\label{eq:rate}
	R_k = \log_2\left(1 + \frac{\left| \Exp\left[\hat{\symsfit{x}}^*_k[\nu] \symsfit{x}_k[\nu]\right] \right|^2}{\Exp\left[|\hat{\symsfit{x}}_k[\nu]|^2\right] - \left| \Exp\left[\hat{\symsfit{x}}^*_k[\nu] \symsfit{x}_k[\nu]\right] \right|^2}\right).
\end{align}
To compute the expected values in \eqref{eq:rate}, the estimate of the transmit signal in \eqref{eq:tx_signal_estimate} can be expanded by using the relation in \eqref{eq:quant_error} and writing the channel as $\symsfit{h}_{mk}[\nu] = \hat{\symsfit{h}}_{mk}[\nu] - \epsilon_{mk}[\nu]$:
\begin{align}\label{eq:estimtate}
	\hat{\symsfit{x}}_{k}[\nu] &= \symsfit{x}_k[\nu] \sqrt{\beta_k P_k} \sum_{m=1}^{M} \Exp\left[\symsfit{w}_{mk}[\nu] \hat{\symsfit{h}}_{mk}[\nu] \right] \notag\\
	&\quad+\symsfit{x}_{k}[\nu] \sqrt{\beta_{k} P_{k}} \sum_{m=1}^{M} \left( \symsfit{w}_{mk}[\nu] \hat{\symsfit{h}}_{mk}[\nu] - \Exp\left[ \symsfit{w}_{mk}[\nu] \hat{\symsfit{h}}_{mk}[\nu]\right] \right) \notag\\
	&\quad+ \sum_{k' \neq k} \symsfit{x}_{k'}[\nu] \sqrt{\beta_{k'} P_{k'}} \sum_{m=1}^{M} \symsfit{w}_{mk}[\nu] \hat{\symsfit{h}}_{mk'}[\nu]\notag\\
	&\quad- \sum_{k'=1}^{K} \symsfit{x}_{k'}[\nu] \sqrt{\beta_{k'} P_{k'}} \sum_{m=1}^{M} \symsfit{w}_{mk}[\nu] \epsilon_{mk}[\nu]\notag\\
	&\quad+ \sum_{m=1}^{M} \symsfit{w}_{mk}[\nu] \symsfit{z}_m[\nu] + \frac{1}{\rho} \sum_{m=1}^{M} \symsfit{w}_{mk}[\nu] \symsfit{e}_m[\nu],
\end{align}
where the Fourier transform $\symsfit{e}_m[\nu]$ of $e_m[n]$ is defined as in \eqref{eq:FT_sig}.  Note that only the first term is correlated to the desired signal.  By assuming that the channel is i.i.d.\ Rayleigh fading, it can be shown that the other terms in \eqref{eq:estimtate}---channel gain uncertainty, interference, channel estimation error, thermal noise, quantization error---are mutually uncorrelated and the variance of each term can be evaluated.  In \cite{mollen2016uplinkArXiv}, for example, it is shown, for one-bit \ADCs, that the variance of the last term asymptotically equals
\begin{align}
	\smash[b]{\Exp\left[\left| \frac{1}{\rho} \sum_{m=1}^{M} \symsfit{w}_{mk}[\nu] \symsfit{e}_m[\nu] \right|^2\right] \to Q, \quad L\to\infty},
\end{align}
if the combiner is normalized such that $\sum_{m=1}^{M} \Exp\left[\left|\symsfit{w}_{mk}[\nu]\right|^2\right] = 1$, which will be assumed here.  This can be generalized to general quantization in a similar way.  The rate in \eqref{eq:rate} can then be written as
\begin{align}\label{eq:rate_quant}
	R_k \to \log_2  \left(1 + \frac{\beta_k P_k c_k G}{\sum_{k'=1}^{K} \beta_{k'} P_{k'}(1 - c_{k'}(1 - I)) + Q + N_0} \right),
\end{align}
as $L \to \infty$, where the \emph{array gain} and \emph{interference} terms are defined as
\begin{align}
	G &\triangleq \left|\sum_{m=1}^{M} \Exp\left[\symsfit{w}_{mk}[\nu]\hat{\symsfit{h}}_{mk}[\nu] \right]\right|^2,\\
	I &\triangleq \smash[b]{\operatorname{Var}\left(\sum_{m=1}^{M} \symsfit{w}_{mk}[\nu] \hat{\symsfit{h}}_{mk'}[\nu]\right)},
\end{align}
where
\begin{align}
	G = \begin{cases}
	M\\
	M-K
	\end{cases}\hspace{-1em},\quad 
	I = \begin{cases}
	1, &\text{for maximum-ratio combining}\\
	0, &\text{for zero-forcing combining}
	\end{cases}.
\end{align}
It is shown in \cite{mollen2016uplinkArXiv} that the limit in \eqref{eq:rate_quant} can approximate the rate with negligible error also for practical delay spreads $L$.  The approximation can even be good for some frequency-flat channels ($L=1$) when the received power $\sum_{k=1}^{K}\beta_kP_k$ is small relative to the noise power $N_0$ or when the number of users is large and there is no dominant user, i.e., no user $k$ for which $\beta_k P_k \gg \sum_{k'\neq k} \beta_{k'} P_{k'}$.  For general frequency-flat channels, however, it is \emph{not} true that the quantization error variance vanishes with increasing number of antennas, as it does for large $L$ in \eqref{eq:rate_quant}; this seems to be overlooked in some of the literature \cite{fan2015uplink, zhang2016spectral, li2016channel, li2016howArXiv}.

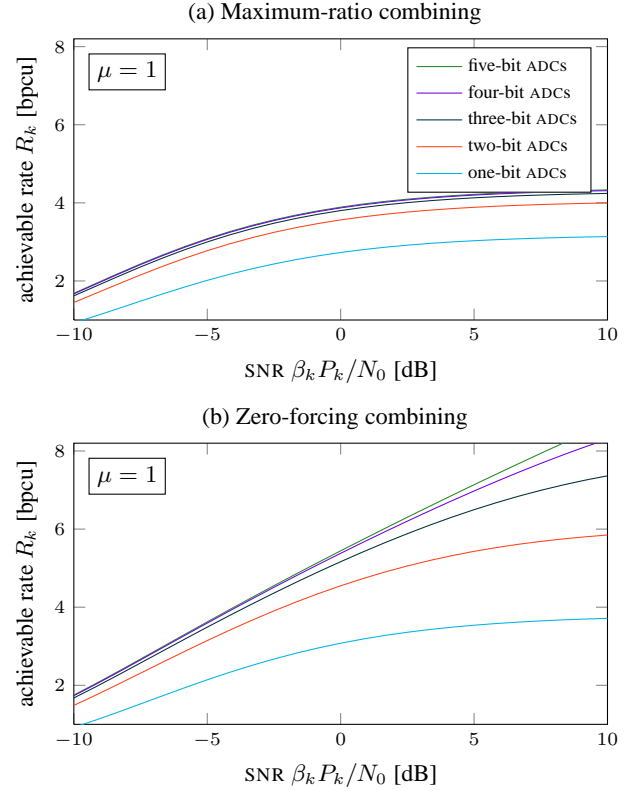
\begin{figure}[t!]
\begin{subfigure}{\linewidth}
	\caption{Maximum-ratio combining}
	\vspace{-1ex}	
	\begin{tikzpicture}
	\def\xmin{-10}
	\def\xmax{10}
	\begin{axis}[
	xmin=\xmin, 
	xmax=\xmax, 
	ymin=1, 
	ymax=8.2, 
	legend pos = north east, 
	height=\figureheight, 
	width=\linewidth,
	xlabel={\SNR $\beta_k P_k / N_0$ [dB]},
	ylabel={achievable rate $R_k$ [bpcu]},
	legend cell align=left,
	]
	
	\def\power{(10^(x/10))}
	\def\nrUsers{5}
	\def\nrAntennae{100}
	\def\pilotExcessFactor{1}
	\def\channelqualityfivebit{(\power * \pilotExcessFactor * \nrUsers / (\power * \pilotExcessFactor * \nrUsers + (\nrUsers * \power + 1) * \fivebiterror + 1))}
	\addplot[color=color5, domain=-10:10] {ln(1 + \power * \channelqualityfivebit * \nrAntennae / (\nrUsers * \power + (\nrUsers * \power + 1) * \fivebiterror + 1))/ln(2)};
	
	\def\channelqualityfourbit{(\power * \pilotExcessFactor * \nrUsers / (\power * \pilotExcessFactor * \nrUsers + (\nrUsers * \power + 1) * \fourbiterror + 1))}
	\addplot[color=color4, domain=\xmin:\xmax] {ln(1 + \power * \channelqualityfourbit * \nrAntennae / (\nrUsers * \power + (\nrUsers * \power + 1) * \fourbiterror + 1))/ln(2)};
	
	\def\channelqualitythreebit{(\power * \pilotExcessFactor * \nrUsers / (\power * \pilotExcessFactor * \nrUsers + (\nrUsers * \power + 1) * \threebiterror + 1))}
	\addplot[color=color3, domain=\xmin:\xmax] {ln(1 + \power * \channelqualitythreebit * \nrAntennae / (\nrUsers * \power + (\nrUsers * \power + 1) * \threebiterror + 1))/ln(2)};
	
	\def\channelqualitytwobit{(\power * \pilotExcessFactor * \nrUsers / (\power * \pilotExcessFactor * \nrUsers + (\nrUsers * \power + 1) * \twobiterror + 1))}
	\addplot[color=color2, domain=\xmin:\xmax] {ln(1 + \power * \channelqualitytwobit * \nrAntennae / (\nrUsers * \power + (\nrUsers * \power + 1) * \twobiterror + 1))/ln(2)};
	
	\def\channelqualityonebit{(\power * \pilotExcessFactor * \nrUsers / (\power * \pilotExcessFactor * \nrUsers + (\nrUsers * \power + 1) * \onebiterror + 1))}
	\addplot[color=color1, domain=\xmin:\xmax] {ln(1 + \power * \channelqualityonebit * \nrAntennae / (\nrUsers * \power + (\nrUsers * \power + 1) * \onebiterror + 1))/ln(2)};
	
	\legend{five-bit \ADCs, four-bit \ADCs, three-bit \ADCs, two-bit \ADCs, one-bit \ADCs}
	
	\node[anchor=north west,draw,outer sep=2mm] at (rel axis cs:0,1) {$\mu = 1$};
	\end{axis}
	\end{tikzpicture}
	\end{subfigure}
	
	\vspace{1ex}
	
	\begin{subfigure}{\linewidth}
	\caption{Zero-forcing combining}
	\vspace{-1ex}
	\begin{tikzpicture}
	\def\xmin{-10}
	\def\xmax{10}
	\begin{axis}[xmin=\xmin, xmax=\xmax, ymin=1, ymax=8.2, height=\figureheight, width=\linewidth,xlabel={\SNR $\beta_k P_k / N_0$ [dB]},ylabel={achievable rate $R_k$ [bpcu]},legend cell align=left]
	
	\def\power{(10^(x/10))}
	\def\nrUsers{5}
	\def\nrAntennae{100}
	\def\pilotExcessFactor{1}
	\def\channelqualityfivebit{(\power * \pilotExcessFactor * \nrUsers / (\power * \pilotExcessFactor * \nrUsers + (\nrUsers * \power + 1) * \fivebiterror + 1))}
	\addplot[color=color5, domain=-10:10] {ln(1 + \power * \channelqualityfivebit * (\nrAntennae - \nrUsers) / (\nrUsers * \power * (1 - \channelqualityfivebit) + (\nrUsers * \power + 1) * \fivebiterror + 1))/ln(2)};
	
	\def\channelqualityfourbit{(\power * \pilotExcessFactor * \nrUsers / (\power * \pilotExcessFactor * \nrUsers + (\nrUsers * \power + 1) * \fourbiterror + 1))}
	\addplot[color=color4, domain=\xmin:\xmax] {ln(1 + \power * \channelqualityfourbit * (\nrAntennae - \nrUsers) / (\nrUsers * \power * (1 - \channelqualityfourbit) + (\nrUsers * \power + 1) * \fourbiterror + 1))/ln(2)};
	
	\def\channelqualitythreebit{(\power * \pilotExcessFactor * \nrUsers / (\power * \pilotExcessFactor * \nrUsers + (\nrUsers * \power + 1) * \threebiterror + 1))}
	\addplot[color=color3, domain=\xmin:\xmax] {ln(1 + \power * \channelqualitythreebit * (\nrAntennae - \nrUsers) / (\nrUsers * \power * (1 - \channelqualitythreebit) + (\nrUsers * \power + 1) * \threebiterror + 1))/ln(2)};
	
	\def\channelqualitytwobit{(\power * \pilotExcessFactor * \nrUsers / (\power * \pilotExcessFactor * \nrUsers + (\nrUsers * \power + 1) * \twobiterror + 1))}
	\addplot[color=color2, domain=\xmin:\xmax] {ln(1 + \power * \channelqualitytwobit * (\nrAntennae - \nrUsers) / (\nrUsers * \power * (1 - \channelqualitytwobit) + (\nrUsers * \power + 1) * \twobiterror + 1))/ln(2)};
	
	\def\channelqualityonebit{(\power * \pilotExcessFactor * \nrUsers / (\power * \pilotExcessFactor * \nrUsers + (\nrUsers * \power + 1) * \onebiterror + 1))}
	\addplot[color=color1, domain=\xmin:\xmax] {ln(1 + \power * \channelqualityonebit * (\nrAntennae - \nrUsers) / (\nrUsers * \power * (1 - \channelqualityonebit) + (\nrUsers * \power + 1) * \onebiterror + 1))/ln(2)};
	
	\node[anchor=north west,draw,outer sep=2mm] at (rel axis cs:0,1) {$\mu = 1$};
	\end{axis}
	\end{tikzpicture}
	\end{subfigure}
	\caption{Rate of a system with 100 antennas and 10 users, where the power is proportional to $1 / \beta_k$ and training is done with $N_\text{p} = KL$ pilots.  The channel is i.i.d.\ Rayleigh fading with uniform power delay profile $h_{mk}[\ell] \sim \CN(0,1 / L)$.  The optimal quantization levels derived in \cite{max1960quantizing} are used.}
	\label{fig:rate}
\end{figure}

The rate $R_k$ is plotted in Figure~\ref{fig:rate} for maximum-ratio and zero-forcing combining.  The transmit powers are allocated proportionally to $1 / \beta_k$ and channel estimation is done with $N_\text{p} = KL$ pilots, i.e., the pilot excess factor $\mu = 1$.  It can be seen that low-resolution \ADCs cause very little performance degradation at spectral efficiencies below \unit[4]{bpcu}.  One-bit \ADCs deliver approximately \unit[40]{\%} lower rates than the equivalent unquantized system and the performance degradation becomes practically negligible with \ADCs with as few as \unit[3]{bit} resolution.  Assuming that the power dissipation in an \ADC is proportional to $2^b$, the use of one-bit \ADCs thus reduces the \ADC power consumption by approximately \unit[6]{dB} at the price of \unit[40]{\%} performance degradation compared to the use of three-bit \ADCs, which deliver almost all the performance of an unquantized system.

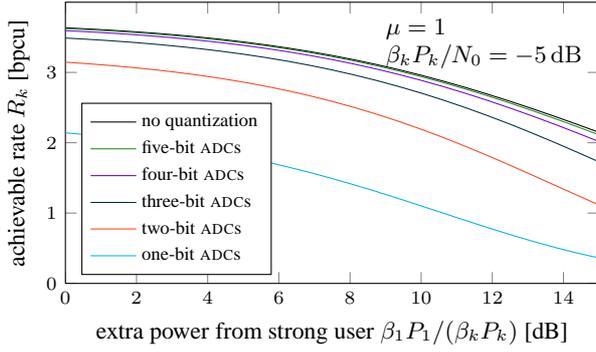
\begin{figure}
		\begin{tikzpicture}[every text node part/.style={align=left}]
		\def\xmin{0}
		\def\xmax{15}
		\begin{axis}[
		xmin=\xmin, 
		xmax=\xmax, 
		ymin=0, 
		legend pos = south west, 
		height=\figureheight, 
		width=\linewidth,
		xlabel={extra power from strong user $\beta_1 P_1 / (\beta_k P_k)$ [dB]},
		ylabel={achievable rate $R_k$ [bpcu]},
		legend cell align=left,
		]
		
		\def\powerdB{-5}
		\def\power{(10^(\powerdB/10))}
		\def\DUpower{(10^((\powerdB + x)/10))}
		\def\nrUsers{5}
		\def\nrAntennae{100}
		\def\pilotExcessFactor{1}
		
		\def\channelquality{(\power * \pilotExcessFactor * \nrUsers / (\power * \pilotExcessFactor * \nrUsers + 1))}
		\addplot[color=black, domain=\xmin:\xmax] {ln(1 + \power * \channelquality * (\nrAntennae - \nrUsers) / (((\nrUsers - 1) * \power + \DUpower) * (1 - \channelquality) + 1))/ln(2)};
		
		\def\channelqualityfivebit{(\power * \pilotExcessFactor * \nrUsers / (\power * \pilotExcessFactor * \nrUsers + ((\nrUsers - 1) * \power + \DUpower + 1) * \fivebiterror + 1))}
		\addplot[color=color5, domain=\xmin:\xmax] {ln(1 + \power * \channelqualityfivebit * (\nrAntennae - \nrUsers) / (((\nrUsers - 1) * \power + \DUpower) * (1 - \channelqualityfivebit) + ((\nrUsers - 1) * \power + \DUpower + 1) * \fivebiterror + 1))/ln(2)};
		
		\def\channelqualityfourbit{(\power * \pilotExcessFactor * \nrUsers / (\power * \pilotExcessFactor * \nrUsers + ((\nrUsers - 1) * \power + \DUpower + 1) * \fourbiterror + 1))}
		\addplot[color=color4, domain=\xmin:\xmax] {ln(1 + \power * \channelqualityfourbit * (\nrAntennae - \nrUsers) / (((\nrUsers - 1) * \power + \DUpower) * (1 - \channelqualityfourbit) + ((\nrUsers - 1) * \power + \DUpower + 1) * \fourbiterror + 1))/ln(2)};
		
		\def\channelqualitythreebit{(\power * \pilotExcessFactor * \nrUsers / (\power * \pilotExcessFactor * \nrUsers + ((\nrUsers - 1) * \power + \DUpower + 1) * \threebiterror + 1))}
		\addplot[color=color3, domain=\xmin:\xmax] {ln(1 + \power * \channelqualitythreebit * (\nrAntennae - \nrUsers) / (((\nrUsers - 1) * \power + \DUpower) * (1 - \channelqualitythreebit) + ((\nrUsers - 1) * \power + \DUpower + 1) * \threebiterror + 1))/ln(2)};
		
		\def\channelqualitytwobit{(\power * \pilotExcessFactor * \nrUsers / (\power * \pilotExcessFactor * \nrUsers + ((\nrUsers - 1) * \power + \DUpower + 1) * \twobiterror + 1))}
		\addplot[color=color2, domain=\xmin:\xmax] {ln(1 + \power * \channelqualitytwobit * (\nrAntennae - \nrUsers) / (((\nrUsers - 1) * \power + \DUpower) * (1 - \channelqualitytwobit) + ((\nrUsers - 1) * \power + \DUpower + 1) * \twobiterror + 1))/ln(2)};
		
		\def\channelqualityonebit{(\power * \pilotExcessFactor * \nrUsers / (\power * \pilotExcessFactor * \nrUsers + ((\nrUsers - 1) * \power + \DUpower + 1) * \onebiterror + 1))}
		\addplot[color=color1, domain=\xmin:\xmax] {ln(1 + \power * \channelqualityonebit * (\nrAntennae - \nrUsers) / (((\nrUsers - 1) * \power + \DUpower) * (1 - \channelqualityonebit) + ((\nrUsers - 1) * \power + \DUpower + 1) * \onebiterror + 1))/ln(2)};
		
		\legend{no quantization, five-bit \ADCs, four-bit \ADCs, three-bit \ADCs, two-bit \ADCs, one-bit \ADCs}
		
		\node[anchor=north east,outer sep=1.2mm] at (rel axis cs:1,1) {$\mu = 1$\\$\beta_k P_k / N_0 = \unit[\powerdB]{dB}$};
		\end{axis}
		\end{tikzpicture}
		\caption{Per-user rate $R_k$ for users $k = 2, \ldots, K$ when user $k = 1$ has a different receive \SNR.  The system has 100 antennas and $K = 10$ users, the channel is i.i.d.\ Rayleigh with uniform power delay profile and is estimated with $N_\text{p} = KL$ pilots.  The optimal quantization levels derived in \cite{max1960quantizing} are used.}
		\label{fig:near_far}
\end{figure}

In \cite{jacobsson2016massive}, it is pointed out that low-resolution \ADCs create a \emph{near--far} problem, where users with relatively weak received power drown in the interference from stronger users.  This is illustrated with a zero-forcing combiner in Figure~\ref{fig:near_far}, where it can be seen how the performance of the weak users degrades if there is a stronger user in the system.  Note that the performance degrades also in the unquantized system, where the imperfect channel estimates prevent perfect suppression of the interference from the strong user.  In the quantized systems, there is a second cause of the performance degradation: With quantization, the pilots are no longer perfectly orthogonal and the quality of the channel estimates is negatively affected by interference from the strong user.  This effect can be seen in \eqref{eq:channel_quality}, where $Q$ scales with the received power $P_\text{rx}$ and thus with the power of the interferer.  

Figure~\ref{fig:near_far}, however, shows that the near--far problem does not become prominent until the received power from the strong user is around \unit[10]{dB} higher than that of the weak users, where the data rate is degraded by approximately \unit[15]{\%} in the unquantized system.  The degradation is larger in the quantized systems but the additional degradation due to quantization is almost negligible when the resolution is \unit[3]{bits} or higher.  With one-bit \ADCs and one strong user with \unit[10]{dB} larger received power, the degradation of the data rate increases to almost \unit[50]{\%}.  Proper power control among users, however, can eliminate the near--far problem altogether.

\section{Conclusion}\label{sec:conclusions}
We have derived an achievable rate for a single-cell massive \MIMO system that takes quantization into account.  The derived rate shows that \ADCs with as low resolution as \unit[3]{bits} can be used with negligible performance loss compared to an unquantized system, also with interference from stronger users.  For example, with three-bit \ADCs, the data rate is decreased by \unit[4]{\%} at spectral efficiencies of \unit[3.5]{bpcu} in a system with 100 antennas that serves 10 users.  It also shows that four-bit \ADCs can be used to accommodate for imperfect automatic gain control---imperfections up to \unit[5]{dB} still result in better performance than the three-bit \ADCs.  One-bit \ADCs can be built from a single comparator and do not need a complex gain control (which \ADCs with more than one-bit resolution need), which simplifies the hardware design of the base station receiver and reduce its power consumption.  The derived rate, however, shows that one-bit \ADCs lead to a significant rate reduction.  For example, one-bit \ADCs lead to a \unit[40]{\%} rate reduction in a system with 100 antennas that serves 10 users at spectral efficiencies of \unit[3.5]{bpcu}.  In the light of the good performance of three-bit \ADCs, whose power consumption should already be small in comparison to other hardware components, the primary reason for the use of one-bit \ADCs would be the simplified hardware design, not the lower power consumption. 

\section{Acknowledgments}
The research leading to these results has received funding from the European Union Seventh Framework Programme under grant agreement number \textsc{ict}-619086 (\textsc{mammoet}), the Swedish Research Council (Vetenskapsrådet) and the National Science Foundation under grant number \textsc{nsf-ccf}-1527079.

\bibliographystyle{IEEEtran}
\bibliography{bib_forkort_namn,bibliografi}
\end{document}